\documentclass{amsart} %{amsart}%{article}%{elsart}
\usepackage{graphicx}
\usepackage{amssymb, amsmath, sidecap}
\usepackage{amsfonts}
\usepackage{amssymb}
\usepackage{float}
\usepackage[version=3]{mhchem}
\usepackage{endnotes}

\newcommand{\R}{\mathbb R}

\def\la{\label}

\def\bt{\begin{thm}}
\def\et{\end{thm}}
\def\bl{\begin{lem}}
\def\el{\end{lem}}
\def\bd{\begin{defi}}
\def\ed{\end{defi}}
\def\bc{\begin{cor}}
\def\ec{\end{cor}}
\def\bp{\begin{proof}}
\def\ep{\end{proof}}
\def\br{\begin{rem}}
\def\er{\end{rem}}

\def\bcon{\begin{conclusion}}
\def\econ{\end{conclusion}}

\newtheorem{thm}{Theorem}[section]
\newtheorem{lem}{Lemma}[section]
\newtheorem{defi}{Definition}[section]

\newtheorem{rem}{Remark}[section]
\newtheorem{cor}{Corollary}[section]

\newtheorem{conclusion}{Physical Conclusion}[section]

\numberwithin{equation}{section}
\numberwithin{figure}{section}

\begin{document}
\title{Phase Transitions for Belousov-Zhabotinsky Reactions}
\author[Ma]{Tian Ma}
\address[TM]{Department of Mathematics, Sichuan University,
Chengdu, P. R. China}

\author[Wang]{Shouhong Wang}
\address[SW]{Department of Mathematics,
Indiana University, Bloomington, IN 47405}
\email{showang@indiana.edu, http://www.indiana.edu/~fluid}

\thanks{The work was supported in part by grants from
the Office of Naval Research,  from the National Science Foundation (US), and from the National Science Foundation of China.}


\begin{abstract}
The main objective of this article is to   study the dynamic phase
transitions associated with the spatial-temporal oscillations of the BZ reactions, given by Field,
K\"or\"os and Noyes, also referred  as the Oregonator. Two criteria are derived to determine 1) existence of either multiple equilibria or spatiotemporal oscillations, and 2) the types of transitions. These criteria gives a complete characterization of the dynamic transitions of the BZ systems from the homogeneous states. The analysis is carried out using a dynamic transition theory developed recently by the authors, which has been successfully applied to a number of problems in science.
\end{abstract}
\maketitle
%\tableofcontents

\section{Introduction}
In 1950's, in his experiments, B. P. Belousov discovered a spatial-temporal
oscillation phenomenon in the concentrations of intermediaries when
citric acid was oxidized by acid bromate in the presence of a cerium
ion catalyst \cite{belousov}.  It is also observed by \cite{zhabo}  that organic
acids and metal ions could  be used as well in the reaction, leading
to spatial-temporal oscillations. It has been considered nowadays that all of the chemical reactions giving rise to oscillations,  and  the actions of catalyst are termed as the Belousov-Zhabotinsky (BZ) reactions. 
BZ reactions are now one of a class of reactions that serve as a classical example of non-equilibrium thermodynamics, resulting in the establishment of a nonlinear chemical oscillator. 

The main objective of this article is to   study the dynamic phase
transitions associated with the spatial-temporal oscillations of the BZ reactions, given by Field,
K\"or\"os and Noyes \cite{FKN}. This  BZ reaction  consists of the
following five irreversible steps:
\begin{equation}
\begin{aligned}
&\ce{A + Y ->[k_1] X},\\
&\ce{X + Y ->[k_2] P},\\
&\ce{B + X ->[k_3] 2X + Z},\\
& \ce{2X ->[k_4] Q},\\
& \ce{ Z ->[k_5] \gamma Y},
\end{aligned}
\label{11.1}
\end{equation}
where $\gamma$ is a stoichiometric factor, $P$ and $Q$ are products which
do not join the reaction again, and
$$X=\text{HBrO}_2,\ \ \ Y=\text{Br}^-,\ \ \ \ Z=\text{Ce}^{4+},\ \ \ \ A=B=\text{BrO}^-_3.$$
It is the temporal oscillation of the curium ion ratio Ce(IV)/Ce(III)
which, with a suitable indicator, is displayed by a color change,
when the reagent is stirred.

The technical method for the analysis is the dynamic transition theory developed recently by the authors \cite{ptd, b-book, chinese-book}.  The main philosophy of the dynamic transition theory is to search for  the full set of  transition states, giving a complete characterization on stability and  transition. The set of transition states is represented by a local attractor. Following this philosophy, the dynamic transition theory is developed  to identify the transition states and to classify them both dynamically and physically. One important ingredient of this theory is the introduction of a dynamic classification scheme of phase transitions. With this classification scheme, phase transitions are classified into three types: Type-I, Type-II and Type-III, which, in more mathematically intuitive terms,  are   called continuous, jump and mixed transitions respectively.  The dynamic transition theory is recently developed by the authors   to identify the transition states and to classify them both dynamically and physically; see above references for details. 
The theory is motivated by phase transition problems in nonlinear sciences. Namely, the mathematical theory is developed under close links to the physics, and in return the theory is applied to the physical problems, although more applications are yet to be explored. With this theory, many long standing phase transition problems are either solved or become more accessible, providing new insights to  both theoretical and experimental studies for the underlying physical problems.

With this method in our disposal, we derive in this article a  characterization of dynamic transitions of the BZ reaction. In particular, the analysis in this article shows  that the  BZ system always undergoes a dynamic transition either to multiple equilibria or to periodic solutions (oscillations), dictated by the sign of a nondimensional computable  parameter $\delta_0-\delta_1$; see (\ref{11.39})  and (\ref{11.43}). 

For the case of transitions to periodic solutions (spatiotemporal oscillations), the Type of transitions (Type-I and Type-II) are determined again by another computable nondimensional parameter. In the multiple equilibrium case, for general domains, the transition is always mixed (Type-III), while for rectangular domain, the transition is either continuous (Type-I) or jump (Type-II) based again on the sign of another nondimensional parameter. To demonstrate the applications, the derived characterization of dynamic transitions of the BZ system is then applied to a special example.

It is worth mentioning that continuous (Type-I) transitions imply that  the concentrations  will stay close to the basic homogeneous state, and the jump (Type-II) transition leads to more drastic changes in the concentrations. The mixed (Type-III)  transition lead to two regions of initial concentrations corresponding to jump and continuous transitions respectively.  In addition, both Type-II and Type-III transitions are accompanied with metastable states, and fluctuations between these metastable states; see  \cite{MW09f} for the related concepts for binary systems.

This article is organized as follows. Section 2  introduces the basic model, and Section 3 study the dynamic transitions of the BZ model, and Section 4 gives an application of the theory and main results.

\section{Field-K\"or\"os-Noyes equations}
The stirred case was considered  by  \cite{FN74}, who derived  a
system of ordinary differential equations for  this reaction.
Here we consider the general cases. Let $X,Y,Z$ be variable, and
$A,B$ be constants. The equations governing (\ref{11.1}) are given by 
\begin{equation}
\begin{aligned}
&\frac{\partial u_1}{\partial t}=\sigma_1\Delta
u_1+k_1au_2-k_2u_1u_2+k_3bu_1-2k_4u^2_1, \\
&\frac{\partial u_2}{\partial t}=\sigma_2\Delta
u_2-k_1au_2-k_2u_1u_2+ \gamma k_5u_3,\\
&\frac{\partial u_3}{\partial t}=\sigma_3\Delta u_3+k_3bu_1-k_5u_3,
\end{aligned}\label{11.2}
\end{equation}
where $u_1,u_2,u_3,a,b$ represent the concentrations of $X,Y,Z,A,$
and $B$, $\sigma_i$  $(i=1,2,3)$  are the diffusivities of $u_i$,    and $k_j$  $(1\leq
j\leq 5)$  are  the reaction coefficients as in (\ref{11.1}). The model
(\ref{11.2}) is also called the Oregonator.

The coefficients $\sigma_i$ and $k_j$ are functions of the
temperature $T$. In fact, $k_j=k^0_je^{-E_i/RT},E_i$ is the
activation energy, and $R$ is  the Boltzmann constant.

The  dimensions of the relevant quantities are  given by:
\begin{align*}
&k_j:\ M^{-1}t^{-1} \text{ for } 1\leq j\leq 4,   &&  k_5:\ t^{-1},\\
& a,b,u_i:\ M \text{ for }  1\leq i\leq 3, &&  \sigma_i:\ l^2t^{-1} \text{ for } 1\leq i\leq 3
\end{align*}
where $t$ is the time, $M$  is  the mole density,   and $l$ is the length. 
Then we introduce  the following nondimensional variables:
\begin{align*}
& u_1=\frac{k_1a}{k_2}u^{\prime}_1, && u_2=\frac{k_3b}{k_2}u^{\prime}_2,\\
& u_3=\frac{k_1k_3}{k_2k_5}abu^{\prime}_3, && t=(k_1k_3ab)^{-{1}/{2}}t^{\prime},\\
&x=Lx^{\prime}, &&\alpha =\left(\frac{k_3b}{k_1a}\right)^{{1}/{2}},\\
&\beta =\frac{2k_1k_4a}{k_2k_3b}, &&\delta
=k_5(k_1k_3ab)^{{1}/{2}}, \\
&\mu_i=\frac{\sigma_i}{l^2(k_1k_2ab)^{{1}/{2}}}&& \text{for } i=1,2,3.
\end{align*}
Omitting the primes, we obtain the following nondimensional form of  (\ref{11.2}):
\begin{equation}
\begin{aligned}
&\frac{\partial u_1}{\partial t}=\mu_1\Delta u_1+\alpha
(u_1+u_2-u_1u_2-\beta u^2_1),\\
&\frac{\partial u_2}{\partial t}=\mu_2\Delta
u_2+\frac{1}{\alpha}(\gamma u_3-u_2-u_1u_2),\\
&\frac{\partial u_3}{\partial t}=\mu_3\Delta u_3+\delta (u_1-u_3),
\end{aligned}
\label{11.3}
\end{equation}
where the unknown functions  are $u_i\geq 0$   $(1\leq i\leq 3)$, and the
parameters are positive constants:
$$\mu_1,\ \mu_2,\ \mu_3,\ \alpha ,\ \beta ,\ \gamma ,\delta >0.$$
Let $\Omega\subset \R^n$, representing the container, be a bounded
domain.

If there is an exchange of materials on the  boundary $\partial\Omega$ to
maintain the level of  concentrations of $X,Y,Z$, then the equations
(\ref{11.3}) are supplemented with the Dirichlet boundary condition
\begin{equation}
u=(u_1,u_2,u_3)=0\ \ \ \ \text{on}\ \partial\Omega .\label{11.4}
\end{equation}

If  there is  no exchange of materials on the boundary, the  equations
are supplemented with the Neumann boundary condition
\begin{equation}
\frac{\partial u}{\partial n}=0\ \ \ \ \text{on}\ \partial\Omega.\label{11.5}
\end{equation}

It is known in  \cite{smoller} that the following region
\begin{equation}
D=\{(u_1,u_2,u_3)\in L^2(\Omega)^3 \ |\ 0<u_i<a_i,\ 1\leq i\leq
3\}\label{11.6}
\end{equation}
is invariant for (\ref{11.3})-(\ref{11.4}), where $a_i$   $(1\leq i\leq
3)$ satisfy
$$a_1>\max \{1,\beta^{-1}\},\ \ \ \ a_2>\gamma a_3,\ \ \ \
a_3>a_1.
$$

The invariant region for  the Neumann boundary condition  case is the same
as that of the Dirichlet boundary condition. For convenience, here
we give the following lemma, which was well known in \cite{smoller, temam88}.
This lemma shows that the model has a global attractor.

\bl\la{l11.2}
 The region $D$ given by (\ref{11.6}) is also
invariant for the problem (\ref{11.3}) with (\ref{11.5}). In
particular this problem possesses a global attractor in $D$.
\el

\bp
 We only need to prove that $D$ is invariant for
(\ref{11.3}) with (\ref{11.5}). It suffices to show that
\begin{equation}
\frac{\partial u}{\partial t}\ \text{points\ inward\ at}\ x\in\partial
D.\label{11.47}
\end{equation}
In fact, by (\ref{11.3}), we see that
\begin{align*}
&\frac{\partial u_1}{\partial t}=\alpha u_2>0 
   &&  \text{as}\ u_1=0,\ u_2>0,\ u_3>0,\\
&\frac{\partial u_2}{\partial t}=\frac{\gamma}{\alpha}u_3>0 
   && \text{as}\ u_2=0,\ u_2>0,\ u_3>0,\\
&\frac{\partial u_3}{\partial t}=\delta u_1>0 
   &&  \text{as}\ u_3=0,\ u_1>0,\ u_2>0,\\
& \frac{\partial u_1}{\partial t}=\alpha [a_1(1-\beta a_1)+(1-a_1)u_2]<0
   &&  \text{as}\ u_1=a_1,\ u_2>0,\\
&\frac{\partial u_2}{\partial t}=-\frac{1}{\alpha} [(a_2-\gamma u_3)+a_2u_1]<0
   && \text{as}\ u_2=a_2,\ u_1>0,\ 0<u_3<a_3,\\
&\frac{\partial u_3}{\partial t}=-\delta (a_3-u_1)<0 
   && \text{as}\ u_3=a_3,\ 0<u_1<a_1,
\end{align*}
where $a_1>\text{max}\{1,\beta^{-1}\},a_2>\gamma a_3$, and $a_3>a_1$.
These properties above show that (\ref{11.47}) is satisfied. The
lemma is proved.
\ep

Let 
\begin{eqnarray*}
&&\R^m_+=\{(x_1,\cdots ,x_m)\in \R^m|\ x_i\geq 0,\ 1\leq i\leq m\},\\
&&\lambda =(\mu_1,\mu_2,\mu_3,\alpha ,\gamma ,\delta )\in \R^6_+.
\end{eqnarray*}
The we define the following function spaces:
\begin{align*}
&H=L^2(\Omega)^3,\\
&H_1=\left\{\begin{aligned} &H^2(\Omega)^3 \cap H^1_0(\Omega)^3 && \text{for boundary condition (\ref{11.4})},\\
&\left\{u\in H^2(\Omega)^3\ |\ \frac{\partial u}{\partial n}=0\ \text{on}\
\partial\Omega \right\} && \text{for boundary condition (\ref{11.5})}.
\end{aligned}
\right.
\end{align*}
Define the operators $L_{\lambda}=A_{\lambda}+B_{\lambda}$ and
$G_{\lambda}: H_1\rightarrow H$ by
\begin{eqnarray*}
&&A_{\lambda}u=\left(\mu_1\Delta u_1,\mu_2\Delta u_2,\mu_3\Delta u_3\right),\\
&&B_{\lambda}u=\left(\alpha u_1+\alpha
u_2,-\frac{1}{\alpha}u_2+\frac{\gamma}{\alpha}u_3,\delta u_1-\delta
u_3\right),\\
&&G(u,\lambda )=\left(-\alpha (u_1u_2+\beta
u^2_1),-\frac{1}{\alpha}u_1u_2,0\right).
\end{eqnarray*}
Thus, the Field-Noyes equations (\ref{11.3}),  with  either  (\ref{11.4}) or
 (\ref{11.5}), take the following operator form:
\begin{align}
& \frac{du}{dt}=L_{\lambda}u+G(u,\lambda ),\label{11.7}
\\
&
u(0)=\varphi\label{11.8}
\end{align}
where $\lambda =(\mu_1,\mu_2,\mu_3,\alpha ,\gamma ,\delta )\in
\R^6_+$.

We note that the solutions $u=(u_1,u_2,u_3)$ of (\ref{11.3}) represent concentrations of chemical materials. Hence only the nonnegative functions $u_i\geq 0$  $(1\leq i\leq 3)$ are
chemically realistic.

\section{Phase Transitions for  BZ Reactions}
\subsection{The model and basic states}
We  consider  case where there is no exchange of materials on the boundary. In this case, the model is supplemented with the Neumann boundary condition (\ref{11.5}), and the system (\ref{11.3})  admits 
two physically realistic constant steady state
solutions:
\begin{equation}
U_0=(0,0,0),\qquad U_1=(u^0_1,u^0_2,u^0_3), 
\label{11.28}
\end{equation}
where 
\begin{align*}
&u^0_1=u^0_3=\sigma ,\qquad u^0_2=\frac{\gamma\sigma}{1+\sigma}=\frac{1}{2}(1+\gamma
-\beta\sigma ),\\
&\sigma =\frac{1}{2\beta}\left[(1-\gamma -\beta )+\sqrt{(1-\gamma
-\beta )^2+4\beta (1+\gamma )}\right].
\end{align*}
It is easy to check that the steady state solution $U_0=0$ is always
unstable, because the linearized equations of (\ref{11.3}) have
three real eigenvalues $\lambda_1,\lambda_2$ and $\lambda_3$
satisfying
$$\lambda_1\lambda_2\lambda_3=\text{det}\left(\begin{array}{ccc}
\alpha&\alpha&0\\
0&-\frac{1}{\alpha}&\frac{\gamma}{\alpha}\\
\delta&0&-\delta
\end{array}
\right)=\delta (\gamma +1)>0,$$ and hence there always at least one
eigenvalue with a positive real part.

Therefore we only need to
consider the transition of (\ref{11.3}) with (\ref{11.5}) at the
steady state solution $U_1$ in (\ref{11.28}). For this purpose, we take the translation
\begin{equation}
u_i=u^{\prime}_i+u^0_i\ \ \ \ (1\leq i\leq 3).\label{11.29}
\end{equation}
Omitting the primes, the problem (\ref{11.3}) with (\ref{11.5})
becomes
\begin{equation}
\begin{aligned}
&\frac{\partial u_1}{\partial t}=\mu_1\Delta u_1+\alpha
[(1-u^0_2-2\beta u^0_1)u_1+(1-u^0_1)u_2-u_1u_2-\beta u^2_1],\\
&\frac{\partial u_2}{\partial t}=\mu_2\Delta
u_2+\frac{1}{\alpha}\left[-u^0_2u_1-(1+u^0_1)u_2+\gamma
u_3-u_1u_2\right],\\
&\frac{\partial u_3}{\partial t}=\mu_3\Delta u_3+\delta (u_1-u_3),\\
&\frac{\partial u}{\partial n}|_{\partial\Omega}=0.
\end{aligned}
\label{11.30}
\end{equation}
Then it suffices to study the phase transition  of (\ref{11.30}) at $u=0$.

\subsection{Linear theory and principle of exchange of stabilities (PES)}
In view of (\ref{11.28}), the linearized eigenvalue equations of (\ref{11.30})
are given by 
\begin{equation}
\begin{aligned}
&\mu_1\Delta u_1-\alpha\left[\frac{1}{2}(\gamma +3\beta\sigma
-1)u_1+(\sigma -1)u_2\right]=\lambda u_1,\\
&\mu_2\Delta u_2-\frac{1}{\alpha}\left[\frac{1}{2}(1+\gamma
-\beta\sigma )u_1+(\sigma +1)u_2-\gamma u_3\right]=\lambda u_2,\\
&\mu_3\Delta u_3+\delta (u_1-u_3)=\lambda u_3,\\
&\frac{\partial u}{\partial n}|_{\partial\Omega}=0.
\end{aligned}
\label{11.31}
\end{equation}

Let $\rho_k$ and $e_k$ be the $k$th eigenvalue and eigenvector of
the Laplace operator $\Delta$ with the Neumann boundary condition:
\begin{equation}
\begin{aligned} &\Delta e_k=-\rho_ke_k,\ \ \ \ (\rho_k\geq 0),\\
&\frac{\partial e_k}{\partial n}|_{\partial\Omega}=0.
\end{aligned}
\label{11.32}
\end{equation}
Let $M_k$ be the matrix given by
\begin{equation}
M_k=\left(\begin{array}{ccc} -\mu_1\rho_k-\frac{\alpha}{2}(\gamma
+3\beta\sigma -1)&-\alpha (\sigma -1)&0\\
-\frac{1}{2\alpha}(1+\gamma -\beta\sigma
)&-\mu_2\rho_k-\frac{1}{\alpha}(\sigma +1)&\frac{\gamma}{\alpha}\\
\delta&0&-\mu_3\rho_k-\delta \end{array}\right).\label{11.33}
\end{equation}
Thus, all eigenvalues $\lambda =\beta_{kj}$ of (\ref{11.31}) satisfy
$$M_kx_{kj}=\beta_{kj}x_{kj},\ \ \ \ 1\leq j\leq 3,\ k=1,2,\cdots
,$$ where $x_{kj}\in \R^3$ is the eigenvector of $M_k$ corresponding
to $\beta_{kj}$. Hence, the eigenvector $u_{kj}$ of (\ref{11.31})
corresponding to $\beta_{kj}$ is
\begin{equation}
u_{kj}(x)=x_{kj}e_k(x),\label{11.34}
\end{equation}
where $e_k(x)$ is as in (\ref{11.32}). In particular, $\rho_1=0$ and
$e_1$ is a constant, and 
\begin{equation}
M_1=\left(\begin{array}{ccc} -\frac{\alpha}{2}(\gamma +3\beta\sigma
-1)&-\alpha (\sigma -1)&0\\
-\frac{1}{2\alpha}(1+\gamma -\beta\sigma )&-\frac{1}{\alpha}(\sigma
+1)&\frac{\gamma}{\alpha}\\
\delta&0&-\delta
\end{array}\right).\label{11.35}
\end{equation}
The eigenvalues $\lambda =\beta_{1j}$  $(1\leq j\leq 3)$ of $M_1$
satisfy
\begin{equation}
\begin{aligned}
&\lambda^3+A\lambda^2+B\lambda +C=0,\\
&A=\delta +\left[\frac{\alpha}{2}(3\beta\sigma +\gamma
-1)+\frac{1}{\alpha}(\sigma +1)\right],\\
&B=2\beta\sigma^2+\gamma -(1-\beta )\sigma
+\delta\left[\frac{\alpha}{2}(3\beta\sigma +\gamma
-1)+\frac{1}{\alpha}(\sigma +1)\right],\\
&C=\delta\sigma (2\beta\sigma +\beta +\gamma -1).
\end{aligned}
\label{11.36}
\end{equation}

It is known that all solutions of (\ref{11.36}) have negative real
parts if and only if
\begin{equation}
A>0,\ \ \ \ C>0,\ \ \ \ AB-C>0.\label{11.37}
\end{equation}

Direct calculation shows that these two parameters $A$ and $C$ in (\ref{11.36}) are positive:
\begin{equation} A>0,\ \ \ \ C>0; \label{11.38}
\end{equation}
see also \cite{HM75}.
Note that
$$\beta_{11}\beta_{12}\beta_{13}=-C<0.$$
It implies that all real eigenvalues of (\ref{11.35}) do not change
their sign, and at least one of these real eigenvalues is negative.

In addition, we  derive,  from $AB-C=0$, the critical number
\begin{equation}
\delta_0=\frac{c+b-a^2+\sqrt{(c+b-a^2)^2+4a^2b}}{2a}, \label{11.39}
\end{equation}
where
\begin{align*}
&a=\frac{\alpha}{2}(3\beta\sigma +\gamma
-1)+\frac{1}{\alpha}(\sigma +1),\\
&b=(1-\beta )\sigma -2\beta\sigma^2-\gamma ,\\
&c=\sigma (2\beta\sigma +\beta +\gamma -1).
\end{align*}
It is then clear that  $\delta_0>0$ if and only if \begin{equation} b=(1-\beta
)\sigma -2\beta\sigma^2-\gamma >0,\label{11.40}
\end{equation}
and under the condition (\ref{11.40})
\begin{equation}
AB-C\left\{\begin{array}{ll}
>0&  \text{ if }  \delta >\delta_0,\\
=0&  \text{ if } \delta =\delta_0,\\
<0&  \text{ if } \delta <\delta_0.
\end{array}
\right.\label{11.41}
\end{equation}

Now we need to check the other eigenvalues $\beta_{kj}$ with $k\geq
2$. By (\ref{11.33}), $\lambda_k=\beta_{kj}$  $(k\geq 2)$ satisfy
\begin{equation}
\lambda^3_k+A_k\lambda^2_k+B_k\lambda_k+C_k=0,\label{11.42}
\end{equation}
where 
\begin{align*}
A_k= & A+(\mu_1+\mu_2+\mu_3)\rho_k, \\
B_k= & B+(\mu_1+\mu_2)\rho_k+a\mu_3\rho_k  +\left(\frac{\sigma
+1}{\alpha}\mu_1+\frac{\alpha}{2}(3\beta\sigma |+\gamma
-1)\mu_2\right)\rho_k  \\
& +(\mu_1\mu_2+\mu_1\mu_3+\mu_2\mu_3)\rho^2_k, \\
C_k= & C+\left[\left(\frac{\sigma
+1}{\alpha}\mu_1+\frac{\alpha}{2}(3\beta\sigma +\gamma
-1)\mu_2\right)\delta -b\mu_3\right]\rho_k \\
& +\left[\mu_1\mu_2\delta +\left(\frac{1+\sigma}{\alpha}\mu_1+
\frac{\alpha}{2}(3\beta\sigma +\gamma
-1)\mu_2\right)\mu_3\right]\rho^2_k +\mu_1\mu_2\mu_3\rho^3_k,
\end{align*}
where $a,b$ are as in (\ref{11.39}), and $A,B,C$  are as in
(\ref{11.36}).

By (\ref{11.28}) it is easy to check that
$$3\beta\sigma +\gamma -1>0.$$
With the condition (\ref{11.40}) we introduce another critical
number
\begin{equation}
\delta_1=\max_{\rho_k\neq
0}\left[\frac{\mu_3\rho_kb-C}{\left(\frac{\sigma
+1}{\alpha}\mu_1+\frac{\alpha}{2}(3\beta\sigma +\gamma
-1)\mu_2\right)\rho_k+\mu_1\mu_2\rho^2_k}-\mu_3\rho_k\right], \label{11.43}
\end{equation}
where $b$ is as in (\ref{11.39}).

Then the following lemma provides characterizes the  PES for  (\ref{11.31}).

\bl\la{l11.1}
Let $\delta_0$ and $\delta_1$ be the numbers
given by (\ref{11.39}) and (\ref{11.43}), and $b$ given by
(\ref{11.39}). When $b<0$, all eigenvalues of (\ref{11.31}) always
have negative real parts, and when $b>0$ the following assertions
hold true:

\begin{itemize}

\item[(1)] Let $\delta_0<\delta_1$ and $k_0\geq 2$ the integer that
$\delta_1$ in (\ref{11.43}) reaches its maximum at $\rho_{k_0}$.
Then $\beta_{k_0l}$ is a real eigenvalue of (\ref{11.31}), and
\begin{eqnarray*}
&&\beta_{k_1}(\delta )\left\{\begin{array}{ll} <0 &\text{ if } \delta
>\delta_1,\\
=0&\text{ if } \delta =\delta_1,\\
>0&\text{ if } \delta <\delta_1,
\end{array}\right.\ \ \ \ \text{for}\ \rho_k=\rho_{k_0}\\
&&\text{Re}\beta_{ij}(\delta_1)<0,\ \ \ \ \forall (i,j)\neq (k,1)\
\text{with}\ \rho_k=\rho_{k_0}.
\end{eqnarray*}

\item[(2)] Let $\delta_0>\delta_1$. Then $\beta_{11}(\delta
)=\bar{\beta}_{12}(\delta )$ are a pair of complex eigenvalues of
(\ref{11.31}) near $\delta =\delta_0$, and
\begin{eqnarray*}
&&\text{Re}\beta_{11}=\text{Re}\beta_{12}\left\{\begin{array}{ll}
<0&\text{ if } \delta >\delta_0,\\
=0&\text{ if } \delta =\delta_0,\\
>0&\text{ if } \delta <\delta_0,
\end{array}
\right.\\
&&\text{Re}\beta_{kj}(\delta_0)<0,\ \ \ \ \forall (k,j)\neq
(1,1),(1,2).
\end{eqnarray*}
\end{itemize}
\el

\bp
In (\ref{11.36}) we see that $A=a+\delta , B=a\delta
-b$. By the direct calculation, we can see that
\begin{equation}
A_k>0,\ \ \ \ A_kB_k-C_k>0,\ \ \ \ \forall k\geq 2.\label{11.44}
\end{equation}

As $b<0$, by (\ref{11.38})-(\ref{11.40}),(\ref{11.42}) and
(\ref{11.44}), for all physically sound parameters
$\mu_1,\mu_2,\mu_3,\alpha ,\beta ,\delta >0$, the following
relations hold true
$$A_i>0,\ \ \ \ C_i>0,\ \ \ \ A_iB_i-C_i>0,\ \ \ \ \forall i\geq
1.$$ Hence, it follows that all eigenvalues $\beta_{kj}(k\geq
1,1\leq j\leq 3)$ of (\ref{11.35}) have negative real parts.

As $b>0$, and $\delta_0<\delta_1$, we infer from (\ref{11.38}) and
(\ref{11.41}) that
\begin{equation}
\text{Re}\beta_{1j}(\delta_1)<0,\ \ \ \ \forall 1\leq j\leq
3.\label{11.45}
\end{equation}
In addition, it is clear that
\begin{equation}
\begin{aligned}
&C_{k_1}\left\{\begin{array}{ll}
 >0 &  \text{ if } \delta >\delta_1,\\
=0&  \text{ if } \delta =\delta_1,\\
<0 &  \text{ if } \delta <\delta_1,
\end{array}
\right.\\
&C_k>0 \qquad  \text{at}\ \delta =\delta_1\ \text{for\ all}\ k\neq k_1.
\end{aligned}
\label{11.46}
\end{equation}
Thus Assertion (1) follows from (\ref{11.44})-(\ref{11.46}).

As $\delta_0>\delta_1$, by (\ref{11.46}) we know that $C_k>0$ at
$\delta =\delta_0$ for all $k\geq 2$. Since the real eigenvalues of
$\beta_{1j}$   $(1\leq j\leq 3)$ are negative, the condition
(\ref{11.41}) implies that there are a pair of complex eigenvalues
$\beta_{11}=\bar{\beta}_{12}$ cross the imaginary axis at $\delta
=\delta_0$. Then Assertions (2) follows. The lemma is proved.
\ep

\subsection{Dynamic Phase Transitions}
By Lemma \ref{l11.1}, as $\delta_1<\delta_0$, the problem (\ref{11.30}) undergoes a dynamic transition 
to a periodic solution from $\delta =\delta_0$. To determine the types of transition, we  
introduce a parameter as follows:
\begin{align}
b_1=&\frac{\rho^2}{D^2E}\big[\frac{3}{D_0}((2D-6-\alpha\gamma\rho^2D_3)F_3-(2\gamma
D_3D_4+\alpha D_5D-6)F_1)  \label{11.48}\\
&-\frac{1}{D_0}\left((\alpha\gamma\rho^2D_3-2D_6)F_1+(2\gamma
D_3D_4)+\alpha D_5D_6)F_3\right)\nonumber\\
&+\frac{1}{D_0}(2\gamma\rho D_3+\alpha\rho D_6+\alpha\gamma\rho
D_3D_5-2\rho^{-1}D_4D_6)F_2\nonumber\\
&+\frac{\alpha}{2D^2E\rho}(2\gamma D_3D_8+\alpha
D_6D_7)(2\rho^{-1}D_6D_8-\alpha\gamma\rho D_3D_7)\nonumber\\
&+\frac{\alpha\rho^2}{2D^2E}(2\gamma D_3+\alpha D_6)(2\gamma
D_3D_8+\alpha D_6D_7)\nonumber\\
&-\frac{\alpha\rho^2}{2D^2E}(2D_6-\alpha\gamma\rho^2D_3)(2\rho^{-2}D_6D_8-\alpha\gamma
D_3D_7)\big],\nonumber
\end{align}
where
\begin{align*}
&\rho =\sqrt{B}=\sqrt{\frac{C}{A}},  
    &&E=\alpha^3(\sigma -1)^3(\delta^2_0+\rho^2)^2,\\
&D^2=\frac{\gamma^2\delta^2_0\rho^2(\sigma
-1)^2+\rho^2((\delta^2_0+\rho^2)^2+\gamma\delta^2_0(\sigma
-1))^2}{N^2(\sigma -1)^2(\delta^2+\rho^2)^4}, 
   &&D_0=\frac{\gamma\delta_0}{\alpha a^2}+\frac{a+2\delta_0}{\alpha
    (\delta -1)},\\
&D_1=\frac{\beta}{\alpha}+\frac{(\alpha A-2)(\beta\sigma +2\beta
+\gamma -1)}{2\alpha (\sigma -1)^2},
   &&D_2=\frac{2-\alpha A}{\alpha^2(\sigma -1)^2},\\
&D_4=A+\alpha\beta\sigma^2+\alpha -3\alpha\beta\sigma -\alpha\beta
-\alpha\gamma ,&&D_3=\delta_0(\sigma -1),\\
&D_5=A+\alpha -\alpha\beta\sigma -2\alpha\beta -\alpha\gamma , 
    &&D_6=(\delta^2_0+\rho^2)^2+\gamma\delta^2_0(\sigma -1),\\
&D_7=1-\beta\sigma -\beta -\gamma , 
    &&D_8=\beta\sigma^2+\beta +1-3\beta\sigma -\gamma ,\\
&F_1=\frac{D_1}{A}-\frac{2\rho^2D_1}{A(A^2+4\rho^2)}-\frac{\rho
D_2}{A^2+4\rho^2},\\
&F_2=\frac{D_2}{A}-\frac{4\rho^2D_2}{A(A^2+4\rho^2)}+\frac{2\rho
D_1}{A^2+4\rho^2},\\
&F_3=\frac{2\rho D_1}{A(A^2+4\rho^2)}+\frac{\rho D_2}{A^2+4\rho^2}
\end{align*}
Here $A,B,C$  are as in (\ref{11.36}), and 
$a$  is as in   (\ref{11.39}).

Then we have the following dynamic  transition theorem.

\bt\la{t11.2}
 Let $\delta_1<\delta_0$, and $b_1$ is the
number given by (\ref{11.48}). Then the problem (\ref{11.30}) undergoes  a transition
to periodic solutions at $\delta =\delta_0$, and the following
assertions hold true:

\begin{itemize}
\item[(1)] When $b_1<0$, the transition is of Type-I, and the system 
bifurcates to a periodic solution on $\delta <\delta_0$ which is an
attractor.

\item[(2)] When $b_1>0$, the transition is of type-II, and the system bifurcates on
$\delta >\delta_0$ to a periodic solution, which is a repeller.
\end{itemize}
\et

\bp 
By Lemma \ref{l11.1}, at $\delta =\delta_0$ there is a pair
of imaginary eigenvalues $\beta_{11}=\bar{\beta}_{12}=-i\rho$ of
(\ref{11.31}). Let $z=\xi +i\eta$ and $z^*=\xi^*+i\eta^*$ be the
eigenvectors and conjugate eigenvectors of (\ref{11.31}) corresponding to
$-i\rho$, i.e. $z$ and $z^*$ satisfy that
\begin{equation}
\left.\begin{array}{l} (M_1+i\rho )z=0,\\
(M^*_1-i\rho )z^*=0,
\end{array}\right.\label{11.49}
\end{equation}
where $M_1$ is the matrix as in (\ref{11.35}),   and  $M^*_1$ the transpose of
$M$. Because $\pm i\rho$ are solutions of (\ref{11.36}), and $AB=C$ at
$\delta =\delta_0$, we deduce that
\begin{equation}
\rho^2=B=C/A.\label{11.49-1}
\end{equation}
For $z=(z_1,z_2,z_3)$, from the first equation of (\ref{11.49}) we get
\begin{equation}
\left.\begin{array}{l} \delta_0z_1=(\delta_0-i\rho )z_3,\\
\alpha (\sigma -1)z_2=\left(-\frac{\alpha}{2}(\gamma +3\beta\sigma
-1)+i\rho\right)z_1.
\end{array}\right.\label{11.49-2}
\end{equation}
Thus, we derive from (\ref{11.49-2}) the eigenvectors $z=\xi +i\eta$ as
follows:
\begin{equation}
\left.\begin{array}{l} \xi
=(\xi_1,\xi_2,\xi_3)=\left(1,-\frac{\gamma +3\beta\sigma
-1}{2(\sigma -1)},\frac{\delta^2_0}{\delta^2_0+\rho^2}\right).\\
\eta =(\eta_1,\eta_2,\eta_3)=\left(0,\frac{\rho}{\alpha (\sigma
-1)},\frac{\rho\delta_0}{\delta^2_0+\rho^2}\right)
\end{array}\right.\label{11.50}
\end{equation}
In the same fashion,  we derive from the second equation of (\ref{11.49}) we derive
the conjugate eigenvectors $z^*=\xi^*+i\eta^*$ as
\begin{equation}
\left.\begin{array}{l}
\xi^*=(\xi^*_1,\xi^*_2,\xi^*_3)=\left(-\frac{\sigma
+1}{\alpha^2(\sigma -1)},1,\frac{\gamma\delta_0}{\alpha
(\delta^2_0+\rho^2)}\right),\\
\eta^*=(\eta^*_1,\eta^*_2,\eta^*_3)=\left(-\frac{\rho}{\alpha
(\sigma -1)},0,-\frac{\gamma\rho}{\alpha
(\delta^2_0+\rho^2)}\right).
\end{array}\right.\label{11.51}
\end{equation}

%As in  (\ref{9.343}), 
It is easy to show that 
\begin{equation}
\begin{array}{l} 
(\xi ,\xi^*) = (\eta,\eta^* )=-\frac{\gamma\rho^2D_3}{H_1},\\
(\xi ,\eta^* )=- (\eta ,\xi^* )=-\frac{\rho D_6}{H_1},
\end{array} \label{11.51-1}
\end{equation}
where $H_1=\alpha (\sigma -1)(\delta^2_0+\rho^2)^2, D_3$ and $D_6$
are as in (\ref{11.48}). It is known that functions $\Psi^*_1+i\Psi^*_2$
given by
\begin{equation}
\left.\begin{array}{l} \Psi^*_1=\frac{1}{(\xi ,\xi^*)}[(\xi
,\xi^*)\xi^*+(\xi ,\eta^*)\eta^*],\\
\Psi^*_2=\frac{1}{(\eta ,\eta^*)}[(\eta ,\xi^*)\xi^*+(\eta
,\eta^*)\eta^*], \end{array}\right.\label{11.51-2}
\end{equation}
also satisfy the second equation of (\ref{11.49}) with
\begin{equation}
\left.\begin{array}{l} (\xi ,\Psi^*_1)=(\eta ,\Psi^*_2)\neq 0,\\
(\xi ,\Psi^*_2)=(\eta ,\Psi^*_1)=0.
\end{array}\right.\label{11.51-3}
\end{equation}

Let $u\in H$ be a solution of (\ref{11.30}) expressed as
$$u=x\xi +y\eta +\Phi (x,y),\ \ \ \ x,y\in \R^1.$$
where $\Phi (x,y)$ is the center manifold function of (\ref{11.30}) at
$\delta =\delta_0$. By (\ref{11.51-3}), the reduced equations of (\ref{11.30})
on the center manifold are given by 
\begin{equation}
\begin{aligned}
&  \frac{dx}{dt}=-\rho y+\frac{1}{(\xi ,\Psi^*_1)}(G(x\xi +y\eta +\Phi) ,\Psi^*_1),\\
& \frac{dy}{dt}=\rho x+\frac{1}{(\eta ,\Psi^*_2)}(G(x\xi +y\eta +\Phi),\Psi^*_2),
\end{aligned}\label{11.52}
\end{equation}
where $G(u)=G(u,u)$ is the bilinear operator defined by
\begin{equation}
G(u,v)=(-\alpha u_1v_2-\alpha\beta
u_1v_1,-\frac{1}{\alpha}u_1v_2,0)\label{11.52-1}
\end{equation}
for $u=(u_1,u_2,u_3),v=(v_1,v_2,v_3)\in H_1$.

We are now in  position to solve the center manifold function $\Phi
(x,y)$. To this end, we need to determine the third eigenvalue
$\beta_{13}$ and eigenvector $\zeta$ of $M_1$ at $\delta =\delta_0$.
We know that
$$\beta_{13}\cdot (i\rho )(-i\rho
)=-\rho^2\beta_{13}=\text{det}M_1=C.$$ By (\ref{11.49-1}) we obtain
$$
\beta_{13}=-A=-(\delta_0+a),
$$
and $a$ is the number as in (\ref{11.39}). Then, from the equation
$$
(M_1-\beta_{13})\zeta =0,
$$
we derive the eigenvector
\begin{equation}
\zeta
=(\zeta_1,\zeta_2,\zeta_3)=\left(1,\frac{A-\frac{\alpha}{2}(\gamma
+3\beta\sigma +1)}{\alpha (\sigma
-1)},-\frac{\delta_0}{a}\right).\label{11.53}
\end{equation}
In the same token, from
$$
(M^*_1-\beta_{13})\zeta^*=0, 
$$
we obtain the conjugate eigenvector as follows:
\begin{equation}
\zeta^*=(\zeta^*_1,\zeta^*_2,\zeta^*_3)=\left(\frac{A-\frac{1}{\alpha}(\sigma
+1)}{\alpha (\sigma -1)},1,-\frac{\gamma}{\alpha
a}\right).\label{11.54}
\end{equation}
On the other hand, from (\ref{11.50}) and (\ref{11.52-1}) it follows that
\begin{equation}
\left.\begin{array}{l} 
G_{11}=G(\xi ,\xi )=(-\alpha (\xi_2+\beta),-\frac{\xi_2}{\alpha},0),\\
G_{12}=G(\xi ,\eta )=(-\alpha\eta_2,-\frac{\eta_2}{\alpha},0),\\
G_{22}=G(\eta ,\eta )=0,\\
G_{21}=G(\eta ,\xi )=0.
\end{array}\right.\label{11.54-1}
\end{equation}
Direct calculation shows that
$$(\zeta ,\zeta^*)=D_0,\ \ \ \ (G_{11},\zeta^*)=D_1,\ \ \ \
(G_{12},\zeta^*)=D_2,$$ and $D_0,D_1,D_2$ are as in (\ref{11.48}).

%By (\ref{1.116-3}), 
By the approximation formula (A.11) in \cite{MW09c}, 
the center manifold function $\Phi$ satisfy
\begin{equation}
\Phi =\Phi_1+\Phi_2+\Phi_3+o(2),\label{11.55}
\end{equation}
where 
\begin{align*}
&{\mathcal{L}}\Phi_1=-x^2P_2G_{11}-xyP_2G_{12}, \\
&({\mathcal{L}}^2+4\rho^2){\mathcal{L}}\Phi_2=2\rho^2(x^2-y^2)P_2G_{11}+4\rho^2xyP_2G_{12}, \\
&({\mathcal{L}}^2+4\rho^2)\Phi_3=\rho (y^2-x^2)P_2G_{12}+2\rho
xyP_2G_{11},
\end{align*}
$P_2:H\rightarrow E_2$ is the canonical projection, 
$E_2=$the orthogonal complement  of $E_1$=span$\{\xi ,\eta )$, and  ${\mathcal{L}}$   is 
the linearized operator of (\ref{11.30}). By %Theorem \ref{t1.21} and 
(\ref{11.34}), it is clear that
\begin{equation}
\left.\begin{array}{l} P_2G_{11}=(G_{11},\zeta^*)\zeta =D_1\zeta ,\\
P_2G_{12}=(G_{12},\zeta^*)\zeta =D_2\zeta .
\end{array}
\right.\label{11.56}
\end{equation}
Hence $\Phi_1,\Phi_2,\Phi_3\in\text{span}\{\zeta\}$. It implies that
\begin{equation}
{\mathcal{L}}\Phi_j=M_1\Phi_j=-A\Phi_j.\label{11.57}
\end{equation}

We infer from (\ref{11.55})-(\ref{11.57})   the center manifold function as follows:
\begin{eqnarray}
\Phi&=&\frac{\zeta}{D_0}\left[\left(\frac{D_1}{A}-\frac{2\rho^2D_1}{A(A^2+4\rho^2)}-\frac{\rho
D_2}{A^2+4\rho^2}\right)x^2\right.\label{11.58}\\
&&+\left(\frac{D_2}{A}-\frac{4\rho^2D_2}{A(A^2+4\rho^2)}+\frac{2\rho
D_1}{A^2+4\rho^2}\right)xy\nonumber\\
&&\left.+\left(\frac{2\rho^2D_1}{A(A^2+4\rho^2)}+\frac{\rho
D_2}{A^2+4\rho^2}\right)y^2\right]+o(2)\nonumber\\
&=&\frac{1}{D_0}(F_1x^2+F_2xy+F_3y^2)\zeta +o(2),\nonumber
\end{eqnarray}
where $F_1,F_2,F_3$ are as in (\ref{11.49}).

Inserting (\ref{11.58}) into (\ref{11.52}),  by (\ref{11.51-2}),  we have
\begin{align}
\frac{dx}{dt}=
&-\rho y+\frac{1}{D^2}
\big[ 
[(\xi,\xi^*)(G_{11},\xi^*)+(\xi,\eta^*)(G_{11},\eta^*)]x^2\label{11.59}\\
&+[(\xi ,\xi^*)(G_{12},\xi^*)+(\xi ,\eta^*)(G_{12},\eta^*)]xy\nonumber\\
&+\frac{1}{D_0}(\xi ,\xi^*)(G(\xi ,\zeta )+G(\zeta ,\xi ),\xi^*) (F_1x^3+F_2x^2y+F_3xy^2)\nonumber\\
&+\frac{1}{D_0}(\xi ,\xi^*)(G(\zeta ,\eta),\xi^*)(F_1x^2y+F_2xy^2+F_3y^3)\nonumber\\
&+\frac{1}{D_0}(\xi ,\eta^*)(G(\xi ,\zeta )+G(\zeta ,\xi),\eta^*) (F_1x^3+F_2x^2y+F_3xy^2)\nonumber\\
&+\frac{1}{D_0}(\xi ,\eta^*)(G(\zeta ,\eta),\eta^*)(F_1x^2y+F_2xy^2+F_3y^3)\big]+o(3),\nonumber\\
\frac{dy}{dt}= &  \rho x+ \frac{1}{D^2}\big[ 
[ (\eta,\xi^*)(G_{11},\xi^*)+(\eta,\eta^*)(G_{11},\eta^*)] x^2\label{11.60}\\
& + [(\eta ,\xi^*)(G_{12},\xi^*)+(\eta,\eta^*)(G_{12},\eta^*)]xy\nonumber\\
&+ \frac{1}{D_0}(\eta ,\xi^*)(G(\xi ,\zeta )+G(\zeta ,\xi),\xi^*) (F_1x^3+F_2x^2y+F_3xy^2)\nonumber\\
&+\frac{1}{D_0}(\eta ,\xi^*)(G(\zeta ,\eta),\xi^*)(F_1x^2y+F_2xy^2+F_3y^3)\nonumber\\
&+\frac{1}{D_0}(\eta ,\eta^*)(G(\xi ,\zeta )+G(\zeta ,\xi),\eta^*)
   (F_1x^3+F_2x^2y+F_3xy^2)\nonumber\\
&+ \frac{1}{D_0}(\eta ,\eta^*)(G(\zeta ,\eta),\eta^*)(F_1x^2y+F_2xy^2+F_3y^3)\big] + o(3),\nonumber
\end{align}
where $D^2=(\xi ,\xi^*)^2+(\xi ,\eta^*)^2$.

Based on (\ref{11.50}), (\ref{11.51}), (\ref{11.52-1})-(\ref{11.54-1}),  we find
\begin{align*}
&(G_{11},\xi^*)=\frac{\alpha D_8}{H_2}, &&  (G_{12},\xi^*)=\frac{2\rho}{H_2},\\
&(G_{11},\eta^*)=\frac{\alpha^2\rho D_7}{2H_2}, && (G_{12},\eta^*)=\frac{\alpha\rho^2}{H_2},\\
&(G(\zeta ,\eta ),\xi^*)=\frac{2\rho}{H_2},&&  (G(\zeta ,\eta),\eta^*)=\frac{\alpha\rho^2}{H_2},\\
&(G(\xi ,\zeta )+G(\zeta ,\xi ),\xi^*)=\frac{2D_4}{H_2},
&&(G(\xi ,\zeta )+G(\zeta ,\xi ),\eta^*)=\frac{\alpha\rho
D_5}{H_2},\\
& H_2=\alpha^2(\sigma -1)^2.
\end{align*}
In view of (\ref{11.51-1}), equations (\ref{11.59}) and (\ref{11.60}) become
\begin{align*}
\frac{dx}{dt}= & -\rho y+\big[a_{20}x^2+a_{11}xy+a_{30}x^3+a_{21}x^2y+a_{12}xy^2+a_{03}y^3\big]
    +o(3),\\
\frac{dy}{dt}= & \rho x+[b_{20}x^2+b_{11}xy+b_{30}x^3+b_{21}x^2y+b_{12}xy^2+b_{03}y^3]+o(3),
\end{align*}
where
\begin{eqnarray*}
&&a_{20}=-\frac{\rho^2}{2H_1H_2D^2}(2\alpha\gamma
D_3D_8+\alpha^2D_6D_7),\\
&&a_{11}=-\frac{\rho^3}{H_1H_2D^2}(2\gamma D_3+\alpha D_6),\\
&&b_{20}=\frac{\rho}{2H_1H_2D^2}(2\alpha
D_6D_8-\gamma\alpha^2\rho^2D_3D_7),\\
&&b_{11}=\frac{\rho^2}{2H_1H_2D^2}(2D_6-\alpha\gamma\rho^2D_3),\\
&&a_{30}=\frac{\rho^2F_1}{D_0H_1H_2D^2}(2\gamma D_3D_4+\alpha
D_5D_6),\\
&&a_{12}=-\frac{\rho^2}{D_0H_1H_2D^2}[(2\gamma D_3D_4+\alpha
D_5D_6)F_3+(2\gamma D_3+\alpha D_6)\rho F_2]\\
&&b_{03}=\frac{\rho^2F_3}{D_0H_1H_2D^2}(2D_6-\alpha\gamma\rho^2D_3),\\
&&b_{21}=\frac{\rho^2}{D_0H_1H_2D^2}[(2\rho^{-1}D_4D_6-\alpha\gamma\rho
D_3D_5)F_2+(2D_6-\alpha\gamma\rho^2D_3)F_1]
\end{eqnarray*}

Due to Theorem~A.3 in \cite{MW10a}, the number
$$
b_1=3(a_{30}+b_{03})+(a_{12}+b_{21})+\frac{2}{\rho}(a_{02}b_{02}-a_{20}b_{20})
+\frac{1}{\rho}(a_{11}a_{20}-b_{11}b_{20})
$$
is the same as that given by (\ref{11.48}). Thus the proof is complete.
\ep

\subsection{Transition to steady state solutions}
Thanks to Lemma \ref{l11.1}, for  $\delta_1>\delta_0$, the transition of
(\ref{11.30}) occurs at $\delta =\delta_1$, which is from real
eigenvalues. Let $\delta_1$ achieve it maximum at
$\rho_{k_0}$   $(k_0\geq 2)$. Assume that $\beta_{kol}(\delta )$ is
simple near $\delta_1$.

\medskip

{\sc First}, we consider the case  where 
\begin{equation}
\int_{\Omega}e^3_{k_0}dx\neq 0.\label{11.64}
\end{equation}
In general, the condition (\ref{11.64}) holds true for the case
where $\Omega\neq (0,L)\times D$,   and $D\subset \R^{n-1}(1\leq n\leq 3)$ is a
bounded open set.

Let 
$$
b_0= [\alpha\beta (\sigma
-1)(\mu_2\rho_{k_0}+\frac{1}{\alpha}(\sigma +1)) -(\alpha\mu_2\rho_{k_0}+2\sigma
)(\mu_1\rho_{k_0}+\frac{\alpha}{2}(\gamma +3\beta\sigma -1))].
$$

\bt\la{t11.4}
 Let the above number $b_0\neq 0$, and
$\delta_1>\delta_0$. Under the condition (\ref{11.64}), the
transition of (\ref{11.30}) at $\delta =\delta_1$ is mixed (Type-III), and the
system bifurcates from $(u,\delta )=(0,\delta_1)$ to a unique steady
state solution $u^{\delta}$ such that $u^{\delta}$ is a saddle on
$\delta >\delta_1$, and an attractor on $\delta <\delta_1$. The
solution $u_{\delta}$ can be expressed as
\begin{equation}
u^{\delta}=C\beta_{k_0 l}(\delta )\xi
e_{k_0}+o(|\beta_{k_0 |}),\label{11.65}
\end{equation}
where $\xi =(\xi_1,\xi_2,\xi_3)$ and constant $C$ are given by
\begin{align*}
&\xi_1=\alpha (\sigma -1)(\mu_3\rho_{k_0}+\delta_1),\\
&\xi_2=-(\mu_1\rho_{k_0}+\frac{\alpha}{2}(\gamma +3\beta\sigma
-1))(\mu_3\rho_{k_0}+\delta_1),\\
&\xi_3=\alpha\delta_1(\sigma -1),\\
&C=\frac{[(\mu_3\rho_{k_0}+\delta_1)^2(\mu_1\rho_{k_0}+\mu_2\rho_{k_0}+\frac{1}{\alpha}
(\sigma +1)+\frac{\alpha}{2}(\gamma +3\beta\sigma -1))-(\sigma
-1)\gamma\delta_1]\int_{\Omega}e_{k_0}dx}{b_0(\mu_3\rho_{k_0}+\delta_1)^3\int_{\Omega}e^3_{k_0}dx}.
\end{align*}
\et

\bp
 We apply Theorem~A.2 in \cite{MW08a} % \ref{t5.9} 
 to prove this theorem. Let $\xi$
and $\xi^*\in \R^3$ be the eigenvectors of $M_{k_0}$ and $M^*_{k_0}$
corresponding to $\beta_{k_0l}(\delta_1)=0$, i.e.
$$M_{k_0}\xi =0,\ \ \ \ M^*_{k_0}\xi^*=0,$$
where $M_{k_0}$ is the matrix (\ref{11.33}) with $k=k_0$. It is easy
to see that $\xi$ is as in (\ref{11.65}), and
\begin{equation}
\begin{aligned}
& \xi^*=(\xi^*_1,\xi^*_2,\xi^*_3), \\
&\xi^*_1=-(\mu_2\rho_{k_0}+\frac{1}{\alpha}(\sigma+1))(\mu_3\rho_{k_0}+\delta_1),\\
&\xi^*_2=\alpha (\mu_3\rho_{k_0}+\delta_1)(\sigma -1),\\
&\xi^*_3=\gamma (\sigma -1).
\end{aligned}
\label{11.66}
\end{equation}
For the operator $G$ in (\ref{11.7}), we can derive that
$$\frac{1}{ (\xi e_{k_0},\xi^*e_{k_0})} (G(y\xi
e_{k_0}),\xi^*e_{k_0} )=-\frac{1}{C}y^2+o(2),
$$ 
where $C$ is as in (\ref{11.65}). Therefore, the theorem follows from Theorem~A.2 in \cite{MW08a}.
% \ref{t5.9}. 
\ep

In the following, we assume that
$$\int_{\Omega}e^3_{k_0}dx=0.$$
Define
\begin{align}
b_1=&\frac{1}{(\xi,\xi^*)}\big[-(\mu_1\rho_{k_0}+\frac{\alpha}{2}(\gamma
+3\beta\sigma
-1))(\alpha\mu_2\rho_{k_0}+2)\int_{\Omega}\phi_1e^2_{k_0}dx   \label{11.67}\\
&  +2\alpha^2\beta (\sigma -1)(\mu_2\rho_{k_0}+\frac{\sigma
+1}{\alpha})\int_{\Omega}\phi_1e^2_{k_0}dx \nonumber \\
& +\alpha (\sigma -1)(\alpha\mu_2\rho_{k_0}+2)\int_{\Omega}\phi_2e^2_{k_0}dx\big],\nonumber
\end{align}
where
\begin{eqnarray*}
(\xi ,\xi^*)&=&\alpha (1-\sigma
)\big[(\mu_1\rho_{k_0}+\mu_2\rho_{k_0}+\frac{\sigma
+1}{\alpha}+\frac{\alpha}{2}(\gamma +3\beta\sigma -1))  \\
&&\ \ \times
(\mu_3\rho_{k_0}+\delta_1)^2-\gamma\delta_1(\sigma -1)\big], 
\end{eqnarray*}
 $\phi =(\phi_1,\phi_2,\phi_3)$ satisfies
\begin{equation}
L\phi =-G(\xi )e^2_{k_0},\label{11.68}
\end{equation}
and  the operators $L$  and $G$   are    defined by
\begin{align*}
&
L\phi =\left\{\begin{aligned}
&\mu_1\Delta\phi_1-\alpha\left[\frac{1}{2}(\gamma +3\beta\sigma
-1)\phi_1+(\sigma -1)\phi_2\right],\\
&\mu_2\Delta\phi_2-\frac{1}{\alpha}\left[\frac{1}{2}(1+\gamma
-\beta\sigma )\mu_1+(\sigma +1)\phi_2\right],\\
&\mu_3\Delta\phi_3+\delta (\phi_1-\phi_3),
\end{aligned}
\right.\\
&
G(\xi )=\left\{\begin{aligned} &\alpha^2(\sigma
-1)(\mu_3\rho_{k_0}+\delta_1)^2\left[\mu_1\rho_{k_0}+\frac{\alpha}{2}(\gamma
+3\beta\sigma -1)-\alpha\beta (\sigma -1)\right],\\
&(\sigma
-1)(\mu_3\rho_{k_0}+\delta_1)^2(\mu_1\rho_{k_0}+\frac{\alpha}{2}(\gamma
+3\beta\sigma -1)), \\
&0.
\end{aligned}
\right.
\end{align*}
Then we have the following theorem.

\bt\la{t11.5}
 Let $b_1\neq 0$ be the number given by
(\ref{11.67}), and $\delta_0<\delta_1$. Then the transition of
(\ref{11.30}) at $\delta =\delta_1$ is continuous (Type-I)  as $b_1<0$, and is
jump (Type-II) as $b_1>0$. Moreover, we have  the following assertions:

\begin{itemize}
\item[(1)] When $b_1>0$, this system bifurcates from $(u,\delta
)=(0,\delta_1)$ to two steady state solutions $u^{\delta}_+$ and
$u^{\delta}_-$ on $\delta >\delta_1$, which are saddles, and no
bifurcation on $\delta <\delta_1$.
\item[(2)] When $b_1<0$, this system bifurcates to two steady state
solutions $u^{\delta}_+$ and $u^{\delta}_-$ on $\delta <\delta_1$
which are attractors, and no bifurcation on $\delta >\delta_1$.
\item[(3)] The bifurcated solutions $u^{\delta}_{\pm}$ can be
expressed as
$$u^{\delta}_{\pm}=\pm\left[-\frac{\int_{\Omega}e^2_{k_0}dx}{(\mu_2\rho_{k_0}+\delta_1)^2b_1}
\beta_{k_0l}(\delta )\right]^{{1}/{2}}\xi
e_{k_0}+o(|\beta_{k_0l}|^{1/2}), 
$$ 
where $\xi\in \R^3$ is as in
(\ref{11.65}), $\beta_{k_0l}(\delta )$ as in Lemma \ref{l11.1}.
\end{itemize}
\et

\bp
 We use Theorem~A.1 in \cite{MW08a} %\ref{t5.8} 
 to verify this theorem. Let
$$
u=y\xi e_{k_0}+\Phi (y),\ \ \ \ y\in \R^1,
$$
and $\Phi (y)$ is the center manifold function. Let $\Phi =y^2\phi
=o(2)$; then by the approximation formula of center manifolds (see (A.10) in \cite{MW09c}), 
%Theorem \ref{t1.36}, 
$\phi$ satisfies
$$L\phi =-G(\xi e_{k_0})=-G(\xi )e^2_{k_0},$$
which is the equation (\ref{11.68}).

We see that
\begin{align*}
(G(y\xi e_{k_0}+\Phi ),\xi^*e_{k_0} )
=&-\int_{\Omega}[\alpha (y\xi_1e_{k_0}+\Phi_1)(y\xi_2e_{k_0}+\Phi_2)\xi^*_1e_{k_0}\\
&  +\alpha\beta (y\xi_1e_{k_0}+\Phi_1)^2\xi^*_1e_{k_0}
+\frac{1}{\alpha}(y\xi_1e_{k_0}+\Phi_1)(y\xi_2e_{k_0}+\Phi_2)\xi^*_2e_{k_0}]dx\\
=&y^3[-(\alpha\xi_2\xi^*_1+2\alpha\beta\xi_1\xi^*_1+\frac{1}{\alpha}\xi_2\xi^*_2)\int_{\Omega}
\phi_1e^2_{k_0}dx\\
& -(\alpha\xi_1\xi^*_1+\frac{1}{\alpha}\xi_1\xi^*_2)\int_{\Omega}\phi_2e^2_{k_0}dx]+o(3)
\end{align*}
Hence, have
$$
\frac{1}{(\xi e_{k_0},\xi^*e_{k_0} ) }(G(y\xi e_{k_0}+\Phi),\xi^*e_{k_0})=
\frac{1}{\int_{\Omega}e^2_{k_0}dx}(\mu_2\rho_{k_0}+\delta_1)^2b_1y^3+o(y^3).
$$
Thus the theorem follows from Theorem~A.1 in \cite{MW08a}.% \ref{t5.8}. 
\ep

\subsection{Stirred case}
Theorem \ref{t11.2} describes the spatial-temporal oscillation for the
BZ reactions of (\ref{11.1}) in a non-stirred condition. 
If the reagent is stirred, the equations (\ref{11.30})
are reduced to the following  system of ordinary differential equations:
\begin{equation}
\begin{aligned}
&\frac{du_1}{dt}=-\alpha \left(\frac{1}{2}(\gamma +3\beta\sigma
-1)u_1+(\sigma -1)u_2-u_1u_2-\beta u^2_1\right),\\
&\frac{du_2}{dt}=-\frac{1}{\alpha}\left(\frac{1}{2}(1+\gamma
-\beta\sigma )u_1+(\sigma +1)u_2-\gamma u_3-u_1u_2\right),\\
&\frac{du_3}{dt}=\delta (u_1-u_3).
\end{aligned}
\label{11.61}
\end{equation}
In this case,  only the transition to periodic
solutions can take place, which is stated in the following
theorem.

\bt\la{t11.3}
 Let $\delta_0>0$ be the number given by
(\ref{11.39}), and $b$ be  as in (\ref{11.48}). Then the system
(\ref{11.61}) undergoes a  dynamic transition to periodic solutions at $\delta
=\delta_0$. Furthermore, the  system bifurcates to an unstable periodic orbit on
$\delta >\delta_0$ as $b>0$, and to a stable periodic orbit on
$\delta <\delta_0$ as $b<0$. In addition,  the bifurcated periodic
solution can be expressed in the following form
\begin{equation}
u_{\delta}=[-b^{-1}\text{Re}\beta_{11}(\delta
)]^{{1}/{2}}(\xi\cos\rho t+\eta\sin\rho
t)+o(|\text{Re}\beta_{11}|^{{1}/{2}}),\label{11.62}
\end{equation}
where $\xi ,\eta$ are as in (\ref{11.54}) and (\ref{11.55}),   and 
$\beta_{11}(\delta )$  is  the first complex eigenvalue as described in
Lemma \ref{l11.1}.
\et

\br\la{r11.2}
{\rm
Since  the constant eigenvector space
$E_1=\text{span}\{u_{11},u_{12},u_{13}\}=\R^3$ is invariant for (\ref{11.30}) where $u_{1j}$ are given by 
(\ref{11.34})), the bifurcated
periodic solution $u_{\lambda}$ in Theorem \ref{t11.2} has the same
equations of (\ref{11.61}) on the center manifold near $\delta
=\delta_0$, given by 
\begin{equation}
\begin{aligned}
&\frac{dx}{dt}=\text{Re}\beta_{11}(\delta )x-\rho (\delta
)y+\frac{1}{(\xi ,\xi^*)}(G(x\xi +y\eta ),\xi^*),\\
&\frac{dy}{dt}=\rho (\delta )x+\text{Re}\beta_{11}(\delta
)x+\frac{1}{(\eta ,\eta^*)}(G(x\xi +y\eta ),\eta^*),
\end{aligned}
\label{11.63}
\end{equation}
where $\xi^*,\eta^*$ are given by (\ref{11.56}) and (\ref{11.57}),
$\rho (\delta )=\text{Im } \beta_{11}(\delta )$. The bifurcated periodic
solution of (\ref{11.61}) is written as
$$u_{\delta}=x(t)\xi +y(t)\eta +o(|x|,|y|).$$
In the polar coordinate system
$$x=r\cos\theta ,\ \ \ \ y=r\sin\theta ,$$
the solution $(x(t),y(t))$ of (\ref{11.63}) can be expressed by
\begin{eqnarray*}
&&x(t)=a(\delta )\cos\rho t+o(|a|),\\
&&y(t)=a(\delta )\sim\rho t+o(|a|).
\end{eqnarray*}
%In the same fashion as used in the proofs of Theorems \ref{t4.13} and \ref{t5.12},
The amplitude $a(\delta )$ satisfies
\begin{eqnarray*}
&&\text{Re}\beta_{11}(\delta )+ba^2(\delta )+o(a^2)=0,\\
&&a(\delta )\rightarrow 0\ \text{as}\ \delta\rightarrow\delta_0,\
a(\delta )>0.
\end{eqnarray*}
Thereby, we get
$$a(\delta )=[-b^{-1}\text{Re}\beta_{11}(\delta
)]^{{1}/{2}}+o(|\text{Re}\beta_{11}|^{{1}/{2}}).
$$ 
Thus, we get the expression (\ref{11.62}).
}\er

\br\la{r11.3}
{\rm
By Lemma \ref{l11.2}, for all physically-sound
parameters, each of the two systems (\ref{11.30}) and (\ref{11.61})  has a
global attractor in the invariant region $D$ in (\ref{11.6}). Hence,
when $b>0$, the bifurcated periodic solution $u_{\delta}$ is a
repeller, and its stable manifold divides $D$ into two disjoint open
sets $D_1$ and $D_2$, i.e., $\bar{D}=\bar{D}_1+\bar{D}_2, D_1\cap
D_2=\emptyset$, such that the equilibrium $U_1=(u^0_1,u^0_2,u^0_3)$
in (\ref{11.28}) attracts $D_1$, and there is another attractor
${\mathcal{A}}_2\subset D_2$ which attracts $D_2.$
}
\er

\section{An example}

We begin with chemical parameters.  In the chemical reaction (\ref{11.1}), the constants given in \cite{FN74} are as follows; see  \cite{HM75}:
\begin{align*}
&k_1=1.34M^{-1}S^{-1},   &&  k_2=1.6\times 10^9M^{-1}S^{-1}, && k_3=8\times 10^3M^{-1}S^{-1},  \\
& k_4=4\times 10^7M^{-1}S^{-1}, && a=b=6\times 10^{-2}M. 
\end{align*}
Both $\gamma$ and $k_5$  are order one parameters, and here we take:
$$\gamma =1,\ \ \ \ k_5=1 \cdot S^{-1},$$
and $a=b$ as a control  parameter. The nondimensional parameters
$\alpha ,\beta  ,\delta$ and $\mu_i$   $(1\leq i\leq 3)$ are given by
\begin{equation}
\begin{aligned}
&\alpha =\left(\frac{k_3b}{k_1a}\right)^{{1}/{2}}=7.727\times
10,   \\
& \beta =\frac{2k_1k_4a}{k_2k_3b}=8.375\times 10^{-6},\\
&\delta =k_5(k_1k_3ab)^{{1}/{2}}=1.035\times 10^2aM^{-1}, \\
   & \mu_i=\frac{1}{(k_1k_2ab)^{\frac{1}{2}}}\frac{\sigma_i}{l^2}=4.664\times
10^{-6}\frac{\sigma_i}{l^2a}M^{-1}S^{-1},
\end{aligned}
\label{11.69}
\end{equation}
for $i=1,2,3$.

\subsection{Stirred case}
In view of (\ref{11.69}),  the numbers in (\ref{11.39})   are as follows:
\begin{equation}
\sigma =7\times 10^2,\ \ \ \ a=9.74,\ \ \ \ b=690.79,\ \ \ \
c=8.21.\label{11.72}
\end{equation}
Hence
\begin{equation}
\delta_0=71.67.\label{11.73}
\end{equation}

Now we need to compute the parameter $b_1$ in (\ref{11.48}). At the
critical value $\delta_0=71.67$, the numbers in (\ref{11.36}) are
\begin{equation}
A=a+\delta_0=81.41,\ \ \ \ B=a\delta_0-b=7.27,\ \ \ \
C=\delta_0c=588.41.\label{11.73-1}
\end{equation}
Then we have
\begin{equation}
\left.\begin{array}{lll} \rho =\sqrt{B}\cong 2.7,&E\cong 4\times
10^{21},&D_0\cong 1.3\times 10^{-2}, \\
F_1\cong 8.3\times 10^{-9},&F_2\cong -2.6\times 10^{-7},&F_3\cong
-8.4\times 10^{-10}, \\
D_3\cong 5\times 10^4,&D_4\cong 4\times 10^2,&D_5\cong 80,\\
D_6\cong 3\times 10^7&D_7\cong -6\times 10^{-3},&D_8\cong 4.
\end{array}\right.\label{11.73-2}
\end{equation}
Thus, by (\ref{11.69}) and (\ref{11.72})-(\ref{11.73-2}), the number $b_1$ in (\ref{11.48})
is given by 
\begin{eqnarray}
b_1&\cong&\frac{\rho^2}{D^2ED_0}\times\left[(\alpha\rho
D_6-2\rho^{-1}D_4D_6)F_2-\alpha
D_5D_6(3F_1-F_3)\right]\label{11.74}\\
&&+\frac{\rho^2}{D^2E}\left[\frac{\alpha^2D^2_6D_7D_8}{ED^2\rho^2}+\frac{\alpha^3\rho^2D^2_6D_7}
{2ED^2}-\frac{3.2\times 10^7D_6D_8}{\rho^2ED^2}\right]\nonumber\\
&\cong&-\frac{\rho^2}{ED_0D^2}\times 4\times
10^3-\frac{3.2D_6D_8}{E^2D^4}\times 10^7\nonumber\\
&&-\frac{6\alpha^2D^2_6D_8}{E^2D^4}\times
10^{-3}-\frac{6\alpha^3\rho^4D^2_6}{2E^2D^4}\times 10^{-3}.\nonumber
\end{eqnarray}
Hence, in the stirred situation, by Theorem \ref{t11.3} and
(\ref{11.73})-(\ref{11.74}), as $\delta <\delta_0=71.67$, the system
(\ref{11.61}) bifurcates from $(u,\delta )=(0,\delta_0)$ to a stable
periodic solution, i.e. the reaction system (\ref{11.1}) undergoes a
temporal oscillation in the concentrations of $X=\text{HBrO}_2,Y=\text{Br}^-$,
and $Z=\text{Ce}^{4+}$.

\subsection{Non-stirred case} 
We consider the problem (\ref{11.30}). In this case, there is
another critical parameter $\delta_1$ defined by (\ref{11.43}). When
$\delta_0<\delta_1$, the system undergoes a  transition to multiple equilibria   at $\delta
=\delta_1$, and when $\delta_0>\delta_1$, the system has  a transition to
periodic solutions at $\delta =\delta_0$. 

By (\ref{11.69}) and (\ref{11.72}), the number $\delta_1$ is given
by
$$\delta_1=\max_{\rho_k}\left[\frac{690.8\mu_3}{9.1\mu_1+0.7\mu_2+\mu_1\mu_2\rho_k}-
\frac{588.4}{(9.1\mu_1+0.7\mu_2)\rho_k+\mu_1\mu_2\rho^2_k}-\mu_3\rho_k\right].$$
It is clear that if
\begin{equation}
\mu_3\leq\mu_1,\mu_2,\label{11.75}
\end{equation}
then in view of (\ref{11.73}) we have
$$\delta_1\leq\frac{690.8\mu_3}{9.1\mu_1+0.7\mu_2}<\frac{690.8}{9.8}<\delta_0.$$

If (\ref{11.75}) is not satisfied, let $\Omega_0$ be a given domain, and $\Omega$  be  the extension or
contraction of $\Omega_0$ from $x_0$ in the scale $L$  $(0<L<\infty )$
defined by
\begin{equation}
\Omega (L,x_0)=\{(x-x_0)L|\ x,x_0\in\Omega_0,x_0\ \text{is
fixed}\}.\label{11.71}
\end{equation}
In this case, the
eigenvalues $\rho_k$  $(k\geq 2)$ of (\ref{11.32}) can be expressed by
$$\rho_k=\frac{C_k}{L^2} \ \ \ \ \text{with}\ C_k>0 \ \ \ \ \text{and}\
C_k\rightarrow\infty\ \text{as}\ k\rightarrow\infty .
$$ 
It follows that there exists an $L_0>0$ such that $\delta_1<\delta_0$ for any
$L<L_0$. Thus, by Theorem \ref{t11.2} and (\ref{11.74}),  we have the
following conclusion.

\bcon\la{pc11.2} 
Under the condition of
(\ref{11.75}) or $\Omega =\Omega (L,x_0)$ with $L<L_0$, the system
(\ref{11.30}) bifurcates from $(u,\delta )=(0,\delta_0)$ to a stable
periodic solution on $\delta <\delta_0$, which implies that the
reaction system (\ref{11.1}) with the Neumann boundary condition
undergoes a temporal oscillation for $\delta <\delta_0$. However,
when $\delta >\delta_0$,  this system is in the trivial state $U_1$ in
(\ref{11.28}).
\econ

{\sc Phase transition at $\delta =\delta_1$.}
By the above discussion, when $\mu_3>\mu_i(i=1,2)$ and $L>L_0$ is
large enough, the condition $\delta_0<\delta_1$ may hold true. In the
following, we assume that $L>L_0$ is sufficiently large, and
$$\mu_1=\mu_2,\ \ \ \ \mu_3=z\mu_1\ \ \ \ (z>1).$$
Then $\delta_1$ becomes
$$\delta_1\simeq\max_{x>0}\left[\frac{691z}{10+x}-\frac{589}{(10+x)x}-xz\right],\
\ \ \ (x=\mu_1\rho_k).
$$ 
It follows that $3<x<4$ for $z>1$. We take
$x=\mu_1\rho_{k_0}=3.5$, then
$$\delta_1=47.7z-12.5$$
Thereby, we deduce  that
$$\delta_1> \delta_0=71.8\ \ \ \ \text{as}\ \frac{\mu_3}{\mu_1}=z>1.8.$$

Now, we investigate the transition of (\ref{11.30}) at $\delta_1$.
For simplicity, we consider the case  where 
$$\Omega =(0,L)\subset \R^1.$$
In this case, the eigenvalues and eigenvectors of (\ref{11.32})
are 
$$\rho_k=(k-1)^2\pi^2/L^2;\ \ \ \ e_k=\cos (k-1)\pi x/L.$$
It is clear that
\begin{align*}
&e^2_{k_0}= (e_1+ e_j)/2   && \text{ with } j=2k_0-1, \\
&\int_{\Omega}e^3_{k_0}dx=0 && \text{ by } k_0\geq 2.
\end{align*}
We need to compute the number $b_1$ in (\ref{11.67}). By
(\ref{11.69}), (\ref{11.72})-(\ref{11.73}), and
$$\mu_3\rho_{k_0}=z\mu_1\rho_{k_0}=z\mu_2\rho_{k_0}=3.5z\ \ \ \
(z>1.8),$$ the vectors $G(\xi )$ and $\phi$ in (\ref{11.68}) are
given by
\begin{eqnarray*}
&&G(\xi )\approx (1.6\times 10^7(\mu_3\rho_{k_0}+\delta_1)^2,3\times
10^3(\mu_3\rho_{k_0}+\delta_1)^2,0),\\
&&\phi =\varphi_0e_1+\varphi_je_j\ (j=2k_0-1),\\
&&M_1\varphi_0=-\frac{1}{2}G(\xi ),\\
&&M_j\varphi_j=-\frac{1}{2}G(\xi ), 
\end{eqnarray*} where
\begin{eqnarray*}
&&M_1=\left(\begin{array}{ccc} -0.7&-5.4\times 10^4&0\\
-1.25\times 10^{-2}&-12.5&1.25\times 10^{-2}\\
\delta_1&0&-\delta_1
\end{array}
\right),\\
&&M_j=\left(\begin{array}{ccc} -14.7&-5.4\times 10^4&0\\
-1.25\times 10^{-2}&-26.5&1.25\times 10^{-2}\\
\delta_1&0&-(\mu_3\rho_j+\delta_1)
\end{array}
\right).
\end{eqnarray*}
The direct calculation shows that
$\text{det}M_1=-8.75\delta_1,\text{det}M_j\simeq -1.5\times
10^4z+4.9\times 10^3$, and
\begin{align*}
& M^{-1}_1=\frac{1}{\text{det}M_1}\left(\begin{array}{ccc}
12.5\delta_1&-5.4\times 10^4\delta_1&*\\
0&0.7\delta_1&*\\
*&*&*
\end{array}
\right), \\
& 
M^{-1}_j=\frac{1}{\text{det}M_j}\left(\begin{array}{ccc}
26.5(\mu_3\rho_j+\delta_1)&-5.4\times 10^4(\mu_3\rho_j+\delta_1)&*\\
-1.25\times 10^{-2}\mu_3\rho_j&14.7(\mu_3\rho_j+\delta_1)&*\\
*&*&*
\end{array}
\right).
\end{align*}
Then we get
\begin{align*}
&\varphi_0=(\varphi_{01},\varphi_{02},\varphi_{03})=-\frac{1}{2}M^{-1}_1G(\xi),
  &&\varphi_{01}=2.5\times 10^6(\mu_3\rho_{k_0}+\delta_1)^2,\\
&\varphi_{02}=1.1\times 10^2(\mu_3\rho_{k_0}+\delta_1)^2,
  &&\varphi_j=(\varphi_{j1},\varphi_{j2},\varphi_{j3})=-\frac{1}{2}M^{-1}_jG(\xi)\\
&\varphi_{j1}=\frac{13.4\times 10^3}{1.5z-0.5}(\mu_3\rho_{k_0}+\delta_1)^2,
&&\varphi_{j2}=\frac{2z+27}{1.5z-0.5}(\mu_3\rho_{k_0}+\delta_1)^2.
\end{align*}
Namely
\begin{align*}
& \phi_1= \varphi_{01}e_1+\varphi_{j_1}e_j 
 = (\mu_3\rho_{k_0}+\delta_1)^2\left[2.5\times
10^6e_1+\frac{13.4\times 10^3}{1.5z-0.5}e_j\right],\\
& \phi_2 = \varphi_{02}e_1+\varphi_{j2}e_j 
  =(\mu_3\rho_{k_0}+\delta_1)^2\left[1.1\times
10^2e_1+\frac{2z+27}{1.5z-0.5}e_j\right].
\end{align*}
Putting $\phi_1$ and $\phi_2$ into (\ref{11.67}),  we derive  that
\begin{eqnarray*}
&&b_1=\frac{(\mu_3\rho_{k_0}+\delta_1)^2L}{(\xi
,\xi^*)}\left(5\times 10^8-\frac{5.7\times
10^6}{1.5z-0.5}-\frac{2z+27}{1.5z-0.5}\right), \\
&&(\xi ,\xi^*)=-699\alpha\left[16.7\times
(\delta_1+3.5z)^2-699\delta_1\right].
\end{eqnarray*}
By $\delta_1>\delta_0=71.8$ and $z>1.8$ we see that $b_1<0$. Hence,
by Theorem \ref{t11.5}, the system (\ref{11.1}) undergoes a dynamic transition  at $\delta
=\delta_1$ to the following steady state solutions
$$u_{\pm}=U_1\pm\left(-\frac{L}{2(\mu_2\rho_{k_0}+\delta_1)^2b_1}\beta_{k_0l}(\delta
)\right)^{{1}/{2}}\xi\cos\frac{(k_0-1)\pi}{L}x+o\left(\beta^{{1}/{2}}_{k_0l}\right),$$
which are attractors.

\bibliographystyle{siam}

\end{document}